\documentclass{article}

\newcommand{\iec}{\mbox{i.\,e.\,}}

\newcommand{\egc}{\mbox{e.\,g.\,}}

\newcommand{\mc}[1]{\ensuremath{\mathcal{#1}}}

\newcommand{\ket}[1]{\ensuremath{\left|  #1 \right\rangle}}
\newcommand{\bra}[1]{\ensuremath{\left\langle #1 \right|}}

\newcommand{\proj}[2]{\ensuremath{\ket{#1} \bra{#2}}}
\newcommand{\tpk}[2]{\ensuremath{\ket{#1}\!\otimes\!\ket{#2}}}

\newcommand{\matel}[3]{\ensuremath{\bra{#1} #2 \ket{#3}}}
 
\newcommand{\op}[1]{\ensuremath{\widehat{\textsf{\ensuremath{#1}}}}}
\newcommand{\opad}[1]{\ensuremath{\op{#1}^{\dagger}}}
\newcommand{\id}{\op{\mathsf{1}}}
\newcommand{\denop}{\ensuremath{\rho}}

\newcommand{\tr}{\textsf{Tr}}

\newcommand{\be}{\begin{equation}}
\newcommand{\ee}{\end{equation}}

\usepackage{chicago}
\usepackage{latexsym}

\renewcommand{\id}[1]{\mathrm{id}_{#1}}
\newtheorem{stage}{Stage}
\newtheorem{stageII}{Stage}

\makeatletter
\renewcommand{\section}{\@startsection
   {section}%
   {1}%
   {0mm}%
   {-\baselineskip}%
   {0.5\baselineskip}%
   {\bfseries\normalsize\centering}}%
\makeatother

\begin{document}

\begin{center}
\LARGE
\textbf{Everettian Rationality: defending Deutsch's approach to probability in the Everett interpretation}

\vspace{0.3cm}

\textbf{\textit{David Wallace$^*$}}

\begin{figure*}[b]
(\textit{December 26, 2002})

* Magdalen College, Oxford University, Oxford OX1 4AU, U.K.

(\textit{e-mail:} david.wallace@magdalen.ox.ac.uk).

\end{figure*}

\normalsize
\end{center}

\vspace{0.7cm}

\begin{quote}
An analysis is made of Deutsch's recent claim to have derived the Born rule
from decision-theoretic assumptions. It is argued that Deutsch's proof
must be understood in the explicit context of the Everett
interpretation, and that in this context, it essentially succeeds. Some
comments are made about the criticism of Deutsch's proof by Barnum, Caves, 
Finkelstein, Fuchs, and Schack; it is argued that the flaw which they
point out in the proof does not apply if the Everett interpretation is
assumed.

A longer version of this paper, entitled \emph{Quantum Probability and Decision Theory,
Revisted}, is available online \cite{webversion}. The present paper will appear in \emph{Studies in the History
and Philosophy of Modern Physics}; confusingly, when it does it will also bear the title
\emph{Quantum Probability and Decision Theory, Revisited}.

\end{quote}

\vspace{0.4cm}

\section{Introduction}\label{intro}

In recent work on the Everett (Many-Worlds) interpretation of quantum
mechanics, it has increasingly been recognized that any version of the
interpretation worth defending will be one in which the basic formalism
of quantum mechanics is left unchanged.  Properties such as the
interpretation of the wave-function as describing a multiverse of
branching worlds, or the ascription of probabilities to the branching
events, must be emergent from the unitary quantum mechanics rather than
added explicitly to the mathematics.  Only in this way is it possible to
save the main virtue of Everett's approach: having an account of quantum
mechanics consistent with the last seventy years of physics, not one in
which the edifice of particle physics must be constructed afresh
\cite[p.\,44]{saundersmetaphysics}.\footnote{This is by no means
\emph{universally} recognized.  Everett-type interpretations can perhaps
be divided into three types: 
\begin{description}
\item[(i)] Old-style ``Many-Worlds''
interpretations in which worlds are added explicitly to the quantum
formalism (see, \egc, \citeN{dewitt} and \citeN{deutsch85}, although
Deutsch has since abandoned this approach; in fact, it is hard to find any remaining
defendants of type (i) approaches). 
\item[(ii)] ``Many-Minds'' approaches in which some intrinsic property of the mind 
is essential to understanding how to reconcile indeterminateness and probability with 
unitary quantum mechanics (see,
\egc, \citeN{albertloewer}, Lockwood \citeyear{lockwood,lockwoodbjps1}, \citeN{donald}, and
\citeN{sudbery}).
\item[(iii)] Decoherence-based approaches, such as those defended by myself 
(Wallace \citeyearNP{wallacestructure,wallaceworlds}), Saunders
\citeyear{saundersdecoherence,saundersmetaphysics,saundersprobability},
Deutsch \citeyear{deutschlockwood,deutschstructure}, Vaidman
\citeyear{vaidman,vaidmanencyclopaedia} and \citeN{zurekprobability}.
\end{description}
For the rest of this paper, whenever I refer to ``the Everett
interpretation'', I shall mean specifically the type (iii) approaches.
This is simply for brevity, and certainly isn't meant to imply anything
about what was intended in Everett's original \citeyear{everett} paper.}

Of the two main problems generally raised with Everett-type
interpretations, the preferred-basis problem looks eminently solvable
without changing the formalism.  The main technical tool towards
achieving this has of course been decoherence theory, which has provided powerful
(albeit perhaps not conclusive) evidence  that the quantum state has a
\textit{de facto} preferred basis and that this basis allows us to
describe the universe in terms of a branching structure of approximately
classical, approximately non-interacting worlds. I have argued elsewhere
(Wallace \citeyearNP{wallacestructure,wallaceworlds}) that there are no purely
conceptual problems with using decoherence to solve the preferred-basis
problem, and that the inexactness of the process should give us no cause
to reject it as insufficient. In particular, the branching events in such a theory 
can be understood,
literally, as replacement of one classical world with several --- so
that in the Schr\"{o}dinger Cat experiment, for instance, after the
splitting there is a part of the quantum state which should be
understood as describing a world in which the cat is alive, and another
which describes a world in which it is dead.  This multiplication comes
about not as a consequence of adding extra, world-defining elements to
the quantum formalism, but as a consequence of an ontology of
macroscopic objects (suggested by \citeNP{realpatterns}) according to 
which they are treated as patterns in the underlying microphysics.

This account applies to human observers as much as to cats: such
an observer, upon measuring an indeterminate event, branches into
multiple observers with each observer seeing a different outcome.  Each
future observer is (initially) virtually a copy of the original
observer, bearing just those causal and structural relations to the
original that future selves bear to past selves in a non-branching
theory.  Since (arguably; see \citeN{parfit} for an extended defence)
the existence of such relations is all that there is to personal
identity, the post-branching observers can legitimately be understood as
future selves of the original observer, and he should care about them
just as he would his unique future self in the absence of branching.

This brings us on to the other main problem with the Everett interpretation, the 
concept of probability. Given that the Everettian description of measurement is a 
deterministic, branching process, how are we to reconcile that with the stochastic
description of measurement used in practical applications of quantum
mechanics?  It has been this problem, as much as the preferred basis
problem, which has led many workers on the Everett interpretation to
introduce explicit extra structure into the mathematics of quantum theory
so as to make sense of the probability of a world as (for instance) a measure over
continuously many identical worlds.  Even some proponents of the 
Many-Minds variant on Everett (notably \citeNP{albertloewer} and Lockwood 
\citeyearNP{lockwood,lockwoodbjps1}), 
who arguably have no difficulty with the preferred-basis problem,
have felt forced to modify quantum mechanics in this way.

It is useful to identify two aspects of the problem.  The first might be
called the \emph{incoherence problem}: how, when every outcome actually
occurs, can it even make sense to view the result of a measurement as 
uncertain?
Even were this solved, there would then remain a \emph{quantitative
problem}: why is that uncertainty quantified according to the quantum probability rule
(\iec, the Born rule),
and not (for instance) some other assignment of probabilities to branches?

Substantial progress has also been made on the incoherence problem. In my view,
the most promising approach is Saunders' `subjective uncertainty' theory of
branching: Saunders argues (via the analogy with Parfittian fission) that an
agent awaiting branching should regard it as \emph{subjectively} indeterministic.
That is, he should expect to become
one future copy or another but not both, and he should be uncertain as
to which he will become. (Saunders' strategy can be found in
\citeN{saundersprobability}, and in the longer version of the current
paper.)
An alternative strategy has been suggested by Vaidman 
\citeyear{vaidman,vaidmanencyclopaedia}:
\emph{immediately} after the branching event (before we actually see the
result of the measurement) the agent knows that he is determinately in
one branch or another but is simply ignorant as to which one.

If progress is being made on the incoherence problem, the quantitative
problem is all the more urgent.
In this context it is extremely interesting that David Deutsch has
claimed \cite{deutschprobability} to derive the quantum probability rule 
from decision theory:
that is, from considerations of pure rationality. It is rather
surprising how little attention his work has received in the
foundational community, though one reason may be that it is very unclear
from his paper that the Everett interpretation is assumed from the
start.\footnote{Nonetheless it \emph{is} assumed: 
\begin{quote}
However, in other respects he [the rational agent] will not behave as if
he believed that stochastic processes occur.  For instance if
\emph{asked} whether they occur he will certainly reply `no', because
the non-probabilistic axioms of quantum theory require the state to
evolve in a continuous and deterministic way.
[\citeNP[pp.\,13]{deutschprobability}; emphasis his.]
\end{quote}
}
If it is tacitly assumed that his work refers instead to some
more orthodox collapse theory, then it is easy to see that the proof is
suspect; this is the basis of the criticisms levelled at Deutsch by
Barnum \textit{et al}, \citeyear{BCFFS}.  Their attack on Deutsch's paper seems to have been
influential in the community; however, it is at best questionable
whether or not it is valid when Everettian assumptions are made
explicit. (This matter will be discussed further below.)

If the Everettian context is made explicit, Deutsch's strategy can be
reconstructed as follows. Assuming that the outcome of a measurement can
in some sense be construed as uncertain (that is, that Saunders', Vaidman's, 
or some other
strategy resolves the incoherence problem), then the `quantitative problem' 
splits into two halves:
\begin{enumerate}
\item What justifies using probabilities to quantify the uncertainty
\emph{at all}?
\item Why use those specific probabilities given by the Born rule?
\end{enumerate}

Fairly obviously, the first of these is not really a quantum-mechanical
problem at all but a more general one --- and one which decision theory
is designed to answer. In decision theory, we start with some general
assumptions about rationality, and deduce that any agent whose
preferences between actions satisfies those assumptions must act as if
they allocated probabilities to possible outcomes and preferred those
actions that maximized expected utility with respect to those
probabilities. Roughly speaking, this is to \emph{define} the
probability assigned to  $X$ by the agent as the shortest odds at which the agent
would be prepared to bet on $X$ occurring.

Deutsch's strategy is to transfer this strategy across
to quantum theory: to start with axioms of rational behavior, apply
them to quantum-mechanical situations, and deduce that rational agents
should quantify their subjective uncertainty in the face of splitting by 
the use of probability. What is striking about the quantum-mechanical
version of decision theory, though, is that rational agents are so
strongly constrained in their behavior that not only must they assign
probabilities to uncertain events, they must assign precisely those
probabilities given by the Born Rule. This discovery might be called
Deutsch's theorem, since it is the central result of Deutsch's paper.

The structure of the paper is as follows. Section \ref{game} gives an
unambiguous definition of Deutsch's quantum games, and derives some
preliminary results about them; section \ref{decision} describes the
decision-theoretic assumptions Deutsch makes. In section
\ref{deutschproof} I run through Deutsch's proof of the Born rule;
section \ref{altproof} gives an alternative proof of my own, from
slightly different assumptions. Sections \ref{barnumI} and
\ref{neutrality} deal with possible criticisms of Deutsch's approach
(section \ref{barnumI} reviews the criticisms made by Barnum \textit{et
al}; section \ref{neutrality} describes a possible problem with the
proof not discussed either by Barnum \textit{et al} or by Deutsch).
Section \ref{conclusion} is the conclusion.

An extended version of this paper \cite{webversion} is available online.

\section{Quantum measurements and quantum games}\label{game}

In this section, I will define Deutsch's notion of a `quantum game' 
--- effectively a bet placed on the outcome of a
measurement. Though I follow Deutsch's definition of a game, my notation
will differ from his in order to resolve some ambiguities in the
definition (first identified by Barnum \emph{et al}, \citeyearNP{BCFFS}).

Informally, a (quantum) game is
to be a three-stage process: a system is prepared in some state; a measurement
of some observable is made on that state; a reward, dependent on the
result of the measurement, is paid to the player. Formally, we will define
a \textbf{game} (in boldface) thus:
\begin{quote}
A \textbf{game} is an ordered triple $\langle
\ket{\psi},\op{X},\mc{P}\rangle$, where:
\begin{itemize}
\item \ket{\psi} is a state in some Hilbert space \mc{H} (technically
the Hilbert space should also be included in the definition, but has
been omitted for brevity);
\item \op{X} is a self-adjoint operator on \mc{H};
\item $\mc{P}$ is a function from the spectrum of \op{X} into the real numbers.
\end{itemize}
\end{quote}

Technically this makes a \textbf{game} into a mathematical object; but obviously we're 
really
interested in physical processes somehow described by that object. We
say that a given process \emph{instantiates} some \textbf{game} $\langle
\ket{\psi},\op{X},\mc{P}\rangle$ if and only if that process consists of:
\begin{enumerate}
\item The preparation of some quantum system, whose state space is
described by \mc{H}, in the state (represented by) \ket{\psi};
\item The measurement, on that system, of the observable (represented by)
\op{X};
\item The provision, in each branch in which result `$x$' was recorded, of
some payment of cash value $\mc{P}(x)$.
\end{enumerate}
We'll define a game (not in boldface) as any process which instantiates a 
\textbf{game}.

The distinction between \textbf{games} and games may seem pedantic:
the whole strategy of mathematical physics is to use mathematical objects to 
represent physical states of affairs, and outside the philosophy of
mathematics there is seldom if ever a need to distinguish between the
two. However, it's crucial to an understanding of Deutsch's proof to notice that the
instantiation relation, between \textbf{games} and games is not
one-to-one. Quite the reverse, in fact: many \textbf{games} can
instantiate a given game (unsurprisingly: there are many ways to
construct a measuring device), and (perhaps more surprisingly) a single
game instantiates many \textbf{games}. We can define an equivalence relation 
$\simeq$ between \textbf{games}:
$\mc{G}\simeq \mc{G}'$ iff $\mc{G}$ and $\mc{G}'$ are instantiated by
the same game. 

To explore the properties of $\simeq$, we need to get precise about what physical 
processes
count as measurements. Since we are working in the Everett framework, we can model
a measurement as follows: let $\mc{H}_s$ be the Hilbert space of some subsystem
of the Universe, and $\mc{H}_e$ be the Hilbert space of the measurement 
device;\footnote{In practice, the
Hilbert space $\mc{H}_e$ would probably have to be expanded to include
an indefinitely large portion of the surrounding environment, since the
latter will inevitably become entangled with the device.}
let \op{X} be the observable to be measured.  

Then a \emph{measurement procedure} for \op{X} is specified by:
\begin{enumerate}
\item Some state \ket{\mc{M}_0} of $\mc{H}_e$, to be interpreted as its
initial (pre-measurement) state; this state must be an element of the
preferred basis picked out by decoherence.
\item Some basis \ket{\lambda_i} of eigenstates of \op{X}, where
$\op{X}\ket{\lambda_i}=x_i \ket{\lambda_i}.$
\item Some set $\{\ket{\mc{M};x_i; \alpha}\}$ of ``readout states'' of 
$\mc{H}_s\otimes \mc{H}_e$,
also elements of the decoherence basis, \emph{at least} one for each 
state \ket{\lambda_a}. 
The states must physically display $x_i$, in
some way measurable by our observer (\egc, by the position of a needle).
\item Some dynamical process, triggered when the device is activated,
and defined by the rule
\be \tpk{\lambda_i}{\mc{M}_0} \longrightarrow \sum_\alpha \mu(\lambda_i;\alpha) 
\ket{\mc{M};x_i;\alpha}\ee
where the $\mu(\lambda_i;\alpha)$ are complex numbers
satisfying $\sum_\alpha |\mu(\lambda_i;\alpha)|^2=1$.
\end{enumerate}

What justifies calling this a `measurement'?  The short answer is that
it is the standard definition; a more principled answer is that the point
of a measurement of \op{X} is to find the value of \op{X}, and that
whenever the value of \op{X} is definite, the measurement process will
successfully return that value.  (Of course, if the value of \op{X} is
not definite then the measurement process will lead to branching of the
device and the observer; but this is inevitable given linearity.)

Observe that:
\begin{enumerate}
\item We are not restricting our attention to so-called 
``non-disturbing'' measurements, in which
$\ket{\mc{M};x_i}=\tpk{\lambda_i}{\mc{M}';x_i}.$  In general
measurements will destroy or at least disrupt the system being measured,
and we allow for this possibility here.
\item The additional label $\alpha$ allows for the fact that many possible
states of the measurement device may correspond to a single measurement
outcome. Even in this case, of course, an observer can predict that whenever 
\ket{\psi} is an eigenstate of \op{X}, all his / her future copies will 
correctly learn 
the value of \op{X}.  In practice most realistic measurements
are likely to be of this form, because the process of magnifying microscopic
data up to the macro level usually involves some random processes.
\item Since a readout state's labelling is a matter not only of physical
facts about that state but also of the labelling conventions used by the
observer, there is no physical difference between a measurement of
\op{X} and one of $f(\op{X})$, where $f$ is an arbitrary one-to-one
function from the spectrum of \op{X} onto some subset of $\Re$: a 
measurement of $f(\op{X})$ may be
interpreted simply as a measurement of $\op{X}$, using a different
labelling convention.  More accurately, there \emph{is} a physical
difference, but it resides in the brain state of the observer (which
presumably encodes the labelling convention in some way) and not in the
measurement device.
\end{enumerate}

To save on repetition, let us now define some general conventions for
\textbf{games}: we will generally use \op{X} for the operator being
measured, and denote its eigenstates by \ket{\lambda_i}; the eigenvalue
of \ket{\lambda_i} will be $x_i$.  (We allow for the possibility of
degenerate \op{X}, so that may have $x_i=x_j$ even though $i \neq j$.)
We write $\sigma(\op{X})$ for the spectrum of \op{X}, and $\op{P}_X(x)$ 
for the projector onto the eigensubspace of
\op{X} with eigenvalue $x$; thus, 
\be \op{X}=\sum_{x\in\sigma(X)}x \,\op{P}_X(x).\ee

For a given \textbf{game} $\mc{G}=\langle \ket{\psi},\op{X},\mc{P}\rangle$, we  also define
the \emph{weight map} $W_\mc{G}:\Re\rightarrow\Re$ by
\be W_\mc{G}(c) = \sum_{x \in 
\mc{P}^{-1}\{c\}}\matel{\psi}{\op{P}_X(x)}{\psi}\ee (that is, the sum ranges over all
$x\in \sigma(\op{X})$ such that $\mc{P}(x)=c$).
It is readily seen that for any game instantiating \mc{G},
$W_\mc{G}(c)$ is the weight of the payoff $c$: that is, the sum of the 
weights of all the branches in which payoff $c$ is given. Because of
this we can refer without confusion to $W_\mc{G}(c)$ as the weight of
$c$.
(Recall
that the \emph{weight} of a branch is simply the squared modulus of the
amplitude of that branch (relative to the pre-branching amplitude, of
course); thus if the state of a measuring device following measurement
is
\be\sum_i \alpha_i \ket{\mc{M};x_i},\ee
then the weight of the branch in which result $x_i$ occurs is
$|\alpha_i|^2$.)

We can now state and prove the:
\\
\begin{quote}
\textbf{Equivalence Theorem}
\begin{enumerate}
\item Payoff Equivalence (PE):
\be \langle \ket{\psi},\op{X},\mc{P}\cdot f\rangle \simeq
\langle \ket{\psi}, f(\op{X}),\mc{P} \rangle \ee
where $f:\sigma(\op{X})\rightarrow\Re$.
\item Measurement Equivalence (ME):
\be \langle \ket{\psi},\op{X},\mc{P}\rangle \simeq 
\langle \op{U}\ket{\psi},\op{X}',\mc{P}'\rangle \ee
where \begin{itemize}
\item \op{U} is a unitary transformation;
\item \op{X} and $\op{X}'$ satisfy $\op{U}\op{X}=\op{X}'\op{U}$;
\item \mc{P} and $\mc{P}'$ agree on $\sigma(\op{X})$.
\end{itemize}
(Note that we allow $\op{U}$ to connect different Hilbert spaces here. If 
$\op{U}$ transforms a fixed Hilbert space, the result simplifies 
to 
\be \langle \ket{\psi},\op{X},\mc{P}\rangle \simeq
\langle \op{U}\ket{\psi},\op{U}\op{X}\opad{U},\mc{P}\rangle .\,\,\,\,\,)\ee
\item General Equivalence (GE):
$\mc{G}\simeq\mc{G}'$  iff  $W_\mc{G}=W_{\mc{G}'}.$

\end{enumerate}
\end{quote}
\textbf{Proof:}
\begin{enumerate}

\item Recall that our definition of a measurement process
involves a set of states \ket{\mc{M};x_i} of the decoherence-preferred basis, 
which are understood as readout states --- and that the rule associating
an eigenvalue $x_i$ with a readout state \ket{\mc{M};x_i} is just a
matter of convention.  Change this convention, then: regard
\ket{\mc{M};x_i} as displaying $f(x_i)$ --- but also change the payoff
scheme: replace a payoff $\mc{P}\cdot f(x)$ upon getting result $x$
with a payoff $\mc{P}(x)$.  These
two changes replace the \textbf{game} 
$\langle \ket{\psi}, \op{X}, \mc{P}\cdot f \rangle$ with 
$\langle \ket{\psi},f(\op{X}), \mc{P} \rangle$  --- but no physical change at all has 
occurred, just a change of labelling convention.  
Hence $\langle \ket{\psi}, \op{X}, \mc{P}\cdot f \rangle \simeq
\langle \ket{\psi},f(\op{X}), \mc{P} \rangle.$

\item For simplicity, let us assume that $\op{X}$ and $\op{X}'$ act on
different Hilbert spaces $\mc{H}$ and $\mc{H}'$. (This assumption can be
relaxed either by a trivial change of the proof, or directly by
realizing \op{U} in two steps, via an auxiliary Hilbert space.)

Because
$\op{U}\op{X}=\op{X}'\op{U}$, it must be possible to label the
eigenstates $\ket{\mu_1}, \ldots \ket{\mu_{n'}}$ of $\op{X}'$ 
so that for $a \leq n$,
$\op{U}\ket{\lambda_i}=\ket{\mu_i}$ and
$\op{X}'\ket{\mu_i}=x_i\ket{\mu_i}.$
Now, without loss of generality take 
$\ket{\psi}=\sum_{i=1}^n \alpha_i \ket{\lambda_i}$, and consider the 
following physical process:
\begin{enumerate}
\item Prepare the system represented by $\mc{H}$ in state \ket{\psi}, 
and the system represented by $\mc{H}'$ in some fixed state \ket{0'}, 
so that the overall quantum state is 
\be \ket{\psi}\otimes \ket{0'}\otimes\ket{\mc{M}_0}\ee
where \ket{\mc{M}_0} is the initial state of some measurement device for
$\mc{H}'$.
\item Operate on $\mc{H}\otimes \mc{H}'$ with some unitary
transformation realizing $\ket{\phi}\otimes\ket{0'}\rightarrow
\ket{0}\otimes \op{U}\ket{\phi}$, where \ket{\phi} is an arbitrary state
of \mc{H} and \ket{0} is some fixed state of \mc{H}. (That such a
transformation exists is trivial.)
\item Discard the system represented by \mc{H}. (This step is just for
notational convenience.) The system retained is now in state $\op{U}\ket{\psi}
\otimes \ket{\mc{M}_0}$.
\item Measure $\op{X}'$ using the following dynamics:
\be \ket{\mu_i}\otimes \ket{\mc{M}_0} \longrightarrow
\ket{\mc{M};x_i;a}\ee
where for each $i$, \ket{\mc{M};x_i;i} is a readout state giving readout 
$x_i$. (The extra $i$ index is only there to allow for degeneracy, and
can be dropped if $\op{X}'$ is non-degenerate.) 
\item The final state is now
\be \sum_{i=1}^n \alpha_i \ket{\mc{M};x_i;i}.\ee
In the branches where result $x_i$ is recorded, give a payoff 
$\mc{P}'(x_i)$.
\end{enumerate}
This process can be described as follows: in steps (a)--(c) we
prepare the state $\op{U}\ket{\psi}$ of $\mc{H}'$, using an auxiliary
system represented by $\mc{H}$. In step (d) we measure the operator $\op{X}'$ on that state, and in
step (e) we provide a payout $\mc{P}'$. This is an instantiation of the \textbf{game} 
$\langle \op{U}\ket{\psi},\op{X}',\mc{P}'\rangle$.

However, suppose we just treat steps (b)--(d) as a black box process.
That process realizes the transformation
\be \left(\sum_{i=1}^n\alpha_i\ket{\lambda_i}\right) \otimes \ket{0'}\otimes
\ket{\mc{M}_0}
\longrightarrow
\sum_{i=1}^n \alpha_i \ket{\mc{M};x_i;i},\ee
which --- by definition of measurement --- is a measurement of \op{X} on
the state \ket{\psi}, using a measurement device with initial state
$\ket{0}\otimes \ket{\mc{M}_0}$.

This observation means that the process (a)--(e) can \emph{also} be described 
in another way: in
step (a) we prepare the state $\ket{\psi}$ of $\mc{H}$; in steps 
(b)--(d) we measure the operator $\op{X}$ on that state (using an
auxiliary system represented by $\mc{H}'$); in step (e) we provide a
payout $\mc{P}$. (Note that the measurement of $\op{U}\ket{\psi}$ gives,
with certainty, some result $x_1, \ldots x_n$, so there is no physical
difference between providing payoff $\mc{P}$ and payoff $\mc{P}'$.)
Thus the process is an instantiation of the \textbf{game} 
$\langle\ket{\psi},\op{X},\mc{P}\rangle$.

There is no physical difference between the two descriptions of (a)--
(e); there is simply a change in how we choose to describe the process.
It follows that $\langle \ket{\psi},\op{X},\mc{P}\rangle\simeq
\langle \op{U}\ket{\psi},\op{X}',\mc{P}'\rangle$.

\item For each $n$, let $\mc{H}^n_0$ be some $n$-dimensional Hilbert space with self-adjoint operator
\op{K}, having eigenstates $\ket{\kappa_1}, \ldots \ket{\kappa_n}$ 
with $\op{K}\ket{\kappa_i}=i \ket{\kappa_i}$. (Technically we should
distinguish between the $\op{K}$ for different $n$, but no ambiguity
will result from this abuse of notation.)

If $\mc{G}=\langle \ket{\psi},\op{X},\mc{P}\rangle$ is any \textbf{game} with $n$ 
distinct payoffs --- that is, elements in the range of \mc{P} --- $c_1, \ldots c_n$ 
with non-zero weights
$w_1, \ldots w_n$ (plus any number of `possible' payoffs with zero
weight), we will show that \mc{G} is equivalent to 
the `canonical' \textbf{game} $\langle \ket{\psi_0},\op{K},\mc{P}_0 \rangle$,
where
\begin{itemize}
\item \ket{\psi_0} is a state in $\mc{H}^n_0$;
\item $\ket{\psi_0}=\sum_{i=1}^n \sqrt{w_i} \ket{\kappa_i}$;
\item $\mc{P}_0(i)=c_i$.
\end{itemize}
This will be sufficient to prove GE. We proceed as follows:
\begin{enumerate}
\item Let \mc{H} be the Hilbert space of \mc{G} (\iec, the Hilbert space
on which \op{X} acts) and let $\mc{S}$ be the direct sum of all eigenspaces of
\op{X} which have nonzero overlap with \ket{\psi}. If \op{U} is the
embedding map of \mc{S} into \mc{H}, and $\mc{P}|_\mc{S}$ is the
restriction of \mc{P} to the spectrum of $\op{X}|_\mc{S}$, then
it follows from ME that 
\be \mc{G} \simeq \langle \ket{\psi}, \op{X}|_\mc{S},
\mc{P}|_\mc{S}\rangle.\ee
We may therefore assume, without loss of generality, that $\mc{S}=\mc{H}$;
that is, that \ket{\psi} has non-zero overlap with all eigenstates of
\op{X}.
\item  Also without loss of generality, we
may assume the eigenstates of \op{X} ordered so that the first $n_1$
give payoff $c_1$, the next $n_2$ give payoff $c_2$, and so on. (We know
each $n_i$ is non-zero, by (a).) Then we can write \ket{\psi} as
\be \ket{\psi}=\sum_{i=1}^N \alpha_i \ket{\lambda_i},\ee 
where each $\alpha_i$ is non-zero.

Now define the normalized vectors $\ket{\mu_1}, \ldots \ket{\mu_n}$ by
\be \ket{\mu_1}= \frac{\alpha_1 \ket{\lambda_1}+ \cdots + 
\alpha_{n_1}\ket{\lambda_{n_1}}}{\sqrt{|\alpha_1|^2 + \cdots +
|\alpha_{n_1}|^2}},\ee
\be \ket{\mu_2}= \frac{\alpha_{n_1+1} \ket{\lambda_{n_1+1}}+ \cdots + 
\alpha_{n_2}\ket{\lambda_{n_2}}}{\sqrt{|\alpha_{n_1+1}|^2 + \cdots +
|\alpha_{n_2}|^2}},\ee
etc.
Then by definition of $W_\mc{G}$, $w_i\equiv  W_\mc{G}(c_i)$, 
we now have \be \ket{\psi}=\sum_{i=1}^n\sqrt{w_i}\ket{\mu_i}.\ee

\item Define $f$ by $f(x)=i$ whenever $x$ is the eigenvalue of an
eigenstate leading to payoff $c_i$. By PE, 
\be \mc{G} \simeq \langle \ket{\psi}, f(\op{X}), \mc{P}\cdot f^{-1}\rangle.\ee
$f(\op{X})$ is an operator which has the $\ket{\mu_i}$ as eigenstates:
$f(\op{X})\ket{\mu_i}=i \ket{\mu_i}.$
 
\item Finally, let $\op{U}$ be a unitary map from $\mc{H}^n_0$ to 
$\mc{H}$, given by $\op{U}\ket{\kappa_i}=\ket{\mu_i}.$
Since $\op{U}\op{K}=f(\op{X}) \op{U}$, we have by ME
\be \mc{G}\simeq\langle \sum_{i=1}^n
\sqrt{w_i}\ket{\kappa_i},\op{K},\mc{P}\cdot f^{-1}\rangle \ee
which is the canonical \textbf{game} above.
\end{enumerate}
\end{enumerate}
Before moving on, it's necessary to cover one further ramification
to his concept of `game' (and `\textbf{game}'): \emph{compound games}. A compound game is
obtained from an existing game by replacing some or all of its
consequences with new games. (For instance, we might measure the spin of
a spin-half particle, and play one of two possible games according to
which spin we obtained.)\footnote{Formally,
\begin{itemize}
\item a \textbf{simple game} is just a \textbf{game} as defined above;
\item a \textbf{compound game} of rank $n$ is a triple $\langle
\ket{\psi},\op{X},\mc{P}\rangle$, where $\mc{P}$ is a map from
$\sigma(\op{X})$ into the set of \textbf{simple games} and \textbf{compound games} 
of rank $n-1$;
\item A compound game is any physical process instantiating a
\textbf{compound game};
\end{itemize}
although in fact we will not ever need to be so formal.}

\section{Decision theory}\label{decision}

To complete our goal of deriving the Born rule, we will need to
introduce some decision-theoretic assumptions about agents' preferences between
games. Following Deutsch, we do so by introducing a \emph{value function}:
a map $\mc{V}$ from the set of \textbf{games}
to the reals, such that if some \textbf{game}'s payoff function is constant and equal to
$c$, then the value of that \textbf{game} is $c$.
(For convenience, we write `$\mc{V}(\ket{\psi},\op{X},\mc{P})$' in place
of `$\mc{V}(\langle\ket{\psi},\op{X},\mc{P}\rangle)$'.)

The idea of the function is that a rational agent prefers
a \textbf{game} $\mc{G}$ to another $\mc{G}'$ just if
$\mc{V}(\mc{G})>\mc{V}(\mc{G}')$. $\mc{V}(\mc{G})$ can be thought of, in
fact, as the `cash value' of $\mc{G}$ to the agent: s/he will be
indifferent between playing \mc{G}, and receiving a reward with a cash value of 
$\mc{V}(\mc{G})$. (It follows that a \textbf{game} whose payoff function is
constant must have $\mc{V}(\mc{G})$ equal to that constant value; hence
the requirement.)

Deutsch now imposes the following restrictions on \mc{V}.

\begin{description}
\item[Dominance:] If $\mc{P}(x)\geq \mc{P}'(x)$ for all $x$, then
\be \mc{V}(\ket{\psi},\op{X},\mc{P})\geq \mc{V}
(\ket{\psi},\op{X},\mc{P}').\ee
\item[Substitutivity:] If $\mc{G}_{comp}$ is a compound \textbf{game} formed from some
\textbf{game} \mc{G} by
substituting for its consequences $c_1, \ldots c_n$ \textbf{games} 
$\mc{G}_1, \ldots \mc{G}_n$ such that $\mc{V}(\mc{G}_i)=c_i$, then 
$\mc{G}_{comp}\simeq \mc{G}$.
\item[Weak additivity:] If $k$ is any real number, then
\be  
\mc{V}( \ket{\psi},\op{X},\mc{P}+k)=
\mc{V}(\ket{\psi},\op{X},\mc{P})+k.\ee
\item[Zero-sum:] For given payoff \mc{P}, let $-\mc{P}$
be defined by $(-\mc{P})(x)=-(\mc{P}(x)).$ Then
\be 
\mc{V}( \ket{\psi},\op{X},-\mc{P})=-
\mc{V}( \ket{\psi},\op{X},\mc{P}).\ee
\end{description}

As with any set of decision-theoretic assumptions, the idea is that any
rational set of preferences between games must be given by some value
function which satisfies these constraints: to violate any one of them
is to be irrational in some way. 

Specifically, \textbf{Dominance} says that if one game invariably leads to better
rewards than another, take the first game. \textbf{Substitutivity} says that, if an
agent is indifferent between getting a definite reward $c$ and playing some game,
s/he should also be indifferent between a chance of getting $c$ and the same chance of
playing that game.

We can motivate \textbf{Weak additivity} like this:  consider any physical process 
which first instantiates 
$\mc{G}=\langle \ket{\psi},\op{X},\mc{P}\rangle$, and then delivers a
reward of value $k$ with certainty. This is physically equivalent to
measuring \ket{\psi} and then receiving, sequentially, two rewards on
getting result $x_a$: one of cash value $\mc{P}(x_a)$ and one of value $k$.
This reward is equivalent to a single one of
value $\mc{P}(x_a)+k$ and so the physical process realizes 
$\langle \ket{\psi},\op{X},\mc{P}+k\rangle$. 

Now, suppose that the fixed reward $k$ is received \emph{before} playing
(the game instantiating) \mc{G}. By \textbf{Substitutivity}, the agent is indifferent
between receiving $k$ then playing \mc{G}, and receiving $k$ then receiving \mc{V}(\mc{G}).
But the latter process is just that of receiving a `lump-sum' payment of $\mc{V}(\mc{G})+k.$

\textbf{Zero-sum} can be motivated as follows: if I
and someone else who shares my exact preferences play some sort of game
in which any gain to one is balanced by a loss to the other, it seems
reasonable to assume that if one of us actively wants to play (that is,
expects to benefit), the other must actively want not to play (that is,
expects to lose out). 

Now suppose $\mc{G}=\langle \ket{\psi},\op{X},\mc{P}\rangle$, and that
I play $\mc{G}'=\langle \ket{\psi},\op{X},\mc{P}-\mc{V}(\mc{G})\rangle$ 
with my alter ego acting as banker; he is thus playing 
$-\mc{G}'=\langle \ket{\psi},\op{X},\mc{V}(\mc{G})-\mc{P}\rangle$.

But by \textbf{Weak additivity}, I am indifferent to playing \mc{G}'
$(\mc{V}(\mc{G}')=0)$. It follows that my alter ego must be indifferent
to playing $-\mc{G}'$, and hence (applying the lemma again) that
\textbf{Zero-sum} holds. (I am grateful to Simon Saunders for this
argument.)

In fact, both \textbf{Weak additivity} and \textbf{Zero-Sum} are special
cases of the following general principle:
\begin{description}
\item[Additivity:] 
\be \mc{V}( \ket{\psi},\op{X},\mc{P}+\mc{P}')
=
\mc{V}(\langle \ket{\psi},\op{X},\mc{P}\rangle)+
\mc{V}(\langle \ket{\psi},\op{X},\mc{P}'\rangle).
\ee
\end{description}

This can be motivated as follows: suppose I know some measurement is to be
carried out, and I want to pay to buy tickets entitling me to a bet on that 
measurement. Each bet is
represented by some payoff function $\mc{P}$, to which I might imagine
assigning a cash value (the largest value I'll pay for the ticket which
allows me to make the bet.) If I assume that the price I'd pay for a given 
ticket doesn't depend on which tickets I've already bought,
\textbf{Additivity} follows.

Deutsch doesn't in fact assume \textbf{Additivity} (though I don't think there's any deep
significance to this) but it will allow us to simplify his proof
considerably. In practice, \textbf{Additivity} is essentially equivalent to the
conjunction of \textbf{Weak additivity}, \textbf{Zero-Sum} and
\textbf{Substitutivity}: the first two allow us to prove
\textbf{Additivity} for games with two possible outcomes, and the third
allows us to build up multi-outcome games from two-outcome ones.

All of these assumptions are essentially independent of quantum mechanics, and
they already allow us to do quite a lot of decision theory: in fact, we can prove 
the
\begin{quote}
\textbf{Probability representation theorem:} If \mc{V} is a value function 
which satisfies  \textbf{Additivity} and \textbf{Dominance}, \mc{V} is given by
\be \label{prob1}\mc{V}(\ket{\psi},\op{X},\mc{P})=
\sum_{x \in\sigma(X)}\Pr_{\psi,X}(x) \mc{P}(x)\ee
where the $\Pr_{\psi,X}(x)$ are real numbers between 0 and 1 which depend
on $\ket{\psi}$ and \op{X} but not on \mc{P}, and where
\be \label{prob2}\sum_{x \in\sigma(X)}\Pr_{\psi,X}(x)=1.\ee
\end{quote}
(The essential idea of the proof is that we \emph{define} the
probability of a measurement outcome as the shortest odds we'd 
accept on its occurrence, and use \textbf{Additivity} to prove this is consistent;
see the appendix for the full proof.)

In fact, this result shows that the decision-theoretic axioms we are
adopting are actually quite strong: they imply, for instance, that it's
rational to bet the mortgage on a one-in-a-million chance of winning the
GNP of Europe. They seem reasonable as long as we restrict our attention
to betting with small sums, however. (And, as I show in \citeN{webversion}, it is possible to improve Deutsch's results by
substantially weakening his decision-theoretic assumptions.)

Nonetheless, the Representation Theorem is still far short of the Born rule. 
No link has been
made between the probabilities $\Pr_{\psi,X}(x)$ and the weight of the
branches, and in fact the Representation Theorem is consistent with
different agents (that is, different value functions) assigning very
different probabilities to the same event.

The connection
to quantum theory comes in entirely through the last assumption made:
\begin{description}
\item[Physicality:]
Two \textbf{games} instantiated by the same physical process have the same value;
that is, $\mc{G}\simeq \mc{G}' \rightarrow \mc{V}(\mc{G})=\mc{V}(\mc{G}').$ 
\end{description}
The motivation for this, obviously, is that real agents have preferences between games, 
not \textbf{games}. I return to this point in section \ref{neutrality}.

\section{Deutsch's proof}\label{deutschproof}

We are now in a position to state and prove

\begin{quote}
\textbf{Deutsch's Theorem:} If $\mc{V}$ is a value function which satisfies 
\textbf{Physicality}, \textbf{Weak additivity}, \textbf{Substitutivity}, 
\textbf{Dominance}, and \textbf{Zero-sum}, then $\mc{V}$ is given uniquely
by the Born rule:
\be \mc{V}(\ket{\psi},\op{X},\mc{P})=\sum_{x\in\sigma(X)}
\matel{\psi}{\op{P}_X(x)}{\psi} \mc{P}(x)
\equiv \sum_{c\in\mc{P}[\sigma(X)]}c W_\mc{G}(c).\ee
\end{quote}

The proof given below follows Deutsch's own proof rather
closely (although some minor changes have been made for clarity or to
conform to my notation and terminology.) In particular, though Deutsch
often uses PE and ME (parts 1 and 2 of the Equivalence Theorem) he
never derives the 3rd part, GE. As such, I make no use of it here
(though see section \ref{altproof}).

As usual, \ket{\lambda_a} will always denote an eigenstate of \op{X}
with some eigenvalue $x_a$. 
It will be convenient, for each operator \op{X}, to define the function
$\id{X}$ as the restriction of the identity map $\id{}(x)=x$ to the
spectrum of $\op{X}$; 
note that 
$\id{f(X)}\cdot f=f \cdot \id{X}.$
Because of PE, if we can prove the theorem for
$\mc{P}=\id{X}$ we can prove it for general $\mc{P}$:
\be \langle\ket{\psi},\op{X},\mc{P}\rangle \simeq 
\langle\ket{\psi},\mc{P}(\op{X}),\id{X}\rangle.\ee
We will therefore take $\id{X}$ as the `default' payoff
function, and will write just $\langle \ket{\psi},\op{X}\rangle$ in
place of $\langle \ket{\psi},\op{X},\id{X}\rangle$. 

\begin{stage} \label{stage1} Let 
$\ket{\psi}=\frac{1}{\sqrt{2}}(\ket{\lambda_1}+\ket{\lambda_2})$.  
Then
$\mc{V}(\ket{\psi},\op{X})=\frac{1}{2}(x_1+x_2)$.
\end{stage}

From \textbf{Weak additivity} and PE, we have
\be\label{24}
\mc{V}(\ket{\psi},\op{X},\id{X})+k=\mc{V}(\ket{\psi},\op{X},\id{X}+k)
=\mc{V}(\ket{\psi},\op{X}+k,\id{X})\ee

Similarly,  \textbf{Zero-Sum} together with another use of PE gives us
\be\label{25}\mc{V}(\ket{\psi},-\op{X})=-\mc{V}(\ket{\psi},\op{X}),\ee
and combining (\ref{24}) and(\ref{25}) gives
\be\label{eq1}\mc{V}(\ket{\psi},-\op{X}+k)=-\mc{V}(\ket{\psi},\op{X})+k.\ee

Now, let $f$ be the function of reflection about the point $1/2
(x_1+x_2)$.  Then
$f(x)=-x+x_1+x_2$.  Provided that \op{X} is non-degenerate and that 
the spectrum of $\op{X}$ is invariant
under the action of $f$, the operator $\op{U}_f$, given by
$\op{U}_f \op{X}\opad{U}_f=f(\op{X})$ is well-defined and leaves
\ket{\psi} invariant. ME then gives us
\be\mc{V}(\ket{\psi},-\op{X}+x_1+x_2)=\mc{V}(\ket{\psi},\op{X}).\ee
Combining this with (\ref{eq1}), we have
\be \mc{V}(\ket{\psi},\op{X})=-\mc{V}(\ket{\psi},\op{X})+x_1+x_2,\ee
which solves to give $\mc{V}(\ket{\psi},\op{X})=\frac{1}{2}(x_1+x_2)$, as
required.

In the general case where $\op{X}$ is degenerate, or has a spectrum which 
is not invariant under the action of $f$, let \mc{S} be the span of 
$\{\ket{\lambda_1},\ket{\lambda_2}\}$ and let
$\mc{V}:\mc{S}\rightarrow\mc{H}$ be the embedding map. ME then gives us
\be \langle \ket{\psi},\op{X}|_\mc{S} \rangle \simeq \langle
\ket{\psi},\op{X} \rangle,\ee 
and the result follows. 

Deutsch refers to this result, with some justice, as `pivotal': it is
the first point in the proof where a connection has been \emph{proved}
between amplitudes and probabilities.
Note the importance in the proof of the symmetry
of \ket{\psi} under reflection, which in turn depends on the equality of
the amplitudes in the superposition; the proof would fail for
$\ket{\psi}=\alpha\ket{\lambda_1}+\beta\ket{\lambda_2},$ unless $\alpha
=\beta$.  

\begin{stage}\label{stage2} If $N=2^n$ for some positive integer $n$, and if
$\ket{\psi}=(1/\sqrt{N})(\ket{\lambda_1}+\cdots+\ket{\lambda_N})$, then
\be \mc{V}(\ket{\psi},\op{X},\mc{P})=(1/N)(x_1+ \cdots + x_N).\ee
\end{stage}
 
The proof is recursive on $n$, and I will give only the first step (the
generalization is obvious).  It relies on the method of forming
composite games, hence on \textbf{Substitutivity}.
Define:
\begin{itemize}
\item $\ket{\psi}=(1/2)(\ket{\lambda_1}+\ket{\lambda_2}
+\ket{\lambda_3}+\ket{\lambda_4})$; 
\item $\ket{A}=(1/\sqrt{2})(\ket{\lambda_1}+\ket{\lambda_2})$;
$\ket{B}=(1/\sqrt{2})(\ket{\lambda_3}+\ket{\lambda_4})$;
\item $y_A=(1/2)(x_1+x_2)$; $y_B=(1/2)(x_3+x_4)$.
\item 
$\op{Y}=y_A \proj{A}{A} + y_B\proj{B}{B}.$
\end{itemize}

Now, the \textbf{game} $\mc{G}=\langle \ket{\psi},\op{Y} \rangle$ has value
$1/4(x_1+x_2+x_3+x_4)$, by Stage \ref{stage1}.
In the $y_A$ branch, a reward of value $1/2(x_1+x_2)$ is given; by
\textbf{Substitutivity} the observer is indifferent between receiving that 
reward
and playing the \textbf{game} $\mc{G}_A=\langle \ket{\psi},\op{X}\rangle$, since the
latter \textbf{game} has the same value.  A similar observation applies in the $y_B$
branch.

So the value to the observer of measuring \op{Y} on \ket{\psi} and then playing either
$G_A$ or $G_B$ according to the result of the measurement is
$1/4(x_1+x_2+x_3+x_4)$.  But the physical process which instantiates
this sequence of \textbf{games} is just
\be   \left(\sum_{i=1}^4\frac{1}{2}\ket{\lambda_i}\right)\otimes
\ket{\mc{M}_0} \rightarrow \sum_{i=1}^4\frac{1}{2}\ket{\mc{M};x_i},\ee
which is also an instantiation of the \textbf{game} $\langle
\ket{\psi},\op{X}\rangle$; hence, the result follows.

\begin{stage}\label{stage3}
Let $N=2^n$ as before, and let $a_1, a_2$ be positive integers
such that $a_1+a_2=N$.  Define \ket{\psi} by 
$\ket{\psi}=\frac{1}{\sqrt{N}}(\sqrt{a_1}
\ket{\lambda_1}+\sqrt{a_2}\ket{\lambda_2})$.
Then 
\be\mc{V}(\ket{\psi})=\frac{1}{N}(a_1 x_1+a_2 x_2).\ee
\end{stage}

Without loss of generality (because of ME) assume \mc{H} is spanned by
\ket{\lambda_1}, \ket{\lambda_2}.
Let $\mc{H}'$ be an $N-$dimensional Hilbert space spanned by states
$\ket{\mu_1},\ldots \ket{\mu_N}$, and define:
\begin{itemize}
\item $\op{Y}=\sum_{i=1}^N i \proj{\mu_i}{\mu_i}$.
\item $f(i)=x_1$ for $i$ between 1 and $a_1$, $f(i)=x_2$ otherwise.
\item $\op{V}:\mc{H}\rightarrow\mc{H}'$ by 
\be\op{V}\ket{\lambda_1}
=\frac{1}{\sqrt{a_1}}\sum_{i=1}^{a_1}\ket{\mu_i} 
\,\,\,\,\mathrm{and}\,\,\,\, 
\op{V}\ket{\lambda_2} = \frac{1}{\sqrt{a_2}}\sum_{i=a_1+1}^{N}\ket{\mu_i}.\ee
\end{itemize}
Then since $f(\op{Y})\op{V}=\op{V}\op{X}$, we have
\be \langle\ket{\psi},\op{X},\mc{P}\rangle \simeq
\langle\op{V}\ket{\psi},f(\op{Y}),\id{f(Y)}\rangle
\simeq \langle \op{V}\ket{\psi},\op{Y},f\cdot\id{Y}.\rangle
\ee
Since in fact $\op{V}\ket{\psi}$ is an equal superposition of all of the
$\ket{\mu_i}$, the result now follows from Stage 2.

Deutsch then goes on to prove the result for arbitrary $N$ (\iec, not 
just $N=2^n$); however, that step can be skipped from the proof without
consequence.

\begin{stage}\label{stage4}
Let $a$ be a positive real number less than 1, and let
$\ket{\psi}=\sqrt{a}\ket{\lambda_1}+\sqrt{1-a}\ket{\lambda_2}$.
Then $\mc{V}(\ket{\psi})=a x_1 + (1-a)x_2$.
\end{stage}

Suppose, without loss of generality, that $x_1 \leq x_2,$ and make the
following definitions:
\begin{itemize}
\item $\mc{G}=\langle \ket{\psi} \rangle$. 
\item $\{a_n\}$ is a decreasing sequence of numbers of form
$a_n=A_n/2^n$, where $A_n$ is a positive integer, and such that
$\lim_{n\rightarrow \infty}a_n = a$.  (This will always be possible, as numbers of this form
are dense in the positive reals.)
\item $\ket{\psi_n}=\sqrt{a_n}\ket{\lambda_1}+\sqrt{1-
a_n}\ket{\lambda_2}$.
\item $\ket{\phi_n}=(1/\sqrt{a_n})(\sqrt{a}\ket{\lambda_1}+\sqrt{a_n-a}\ket{\lambda_2}.$ 
\item $\mc{G}_n = \langle\ket{\psi_n}\rangle$.
\item $\mc{G}_n'= \langle\ket{\phi_n} \rangle$.
\end{itemize}

Now, from Stage \ref{stage3} we know that $\mc{V}(\mc{G}_n)=a_n x_1 + (1-a_n) x_2.$
We don't know the value of $\mc{G}_n'$, but by \textbf{Dominance} we know that it is 
at least $x_1$.  Then, by \textbf{Substitutivity}, the value to the observer of measuring \ket{\psi_n}, 
then receiving $x_2$ euros if the result is $x_2$ and playing
$\mc{G}_n'$ if the result is $x_1$, is at least as great as the
$\mc{V}(\mc{G}_n).$

But this sequence of games is just an
instantiation of \mc{G}, for its end state is one in which a reward of
$x_1$ euros is given with amplitude ${a}$ and a reward of $x_2$ euros with
amplitude $\sqrt{1-a}$.  It follows that $\mc{V}(\mc{G})\geq
\mc{V}(\mc{G}_n)$ for all $n$, and hence that $\mc{V}(\mc{G})\geq a x_1
+ (1-a) x_2.$   

A similar argument with an increasing sequence establishes that
$\mc{V}(\mc{G})\leq a x_1+ (1-a) x_2$, and the result is proved.

\begin{stage}
Let $\alpha_1, \alpha_2$ be complex numbers such that
$|\alpha_1|^2+|\alpha_2|^2=1$,
and let $\ket{\psi}=\alpha_1\ket{\lambda_1}+\alpha_2\ket{\lambda_2}$.
Then $\mc{V}(\ket{\psi})=|\alpha_1|^2 x_1 + |\alpha_2|^2 x_2$.
\end{stage}

This is an immediate consequence of ME and Stage \ref{stage4}: 
let $\op{U}=\sum_i \exp(i \theta_i)
\proj{\lambda_i}{\lambda_i}$; then \op{U} leaves \op{X} invariant
and so $\langle \op{U}\ket{\psi},\op{X}\rangle \simeq
\ket{\psi},\op{X}\rangle;$ but the eigenstate $\op{U}\ket{\psi}$ has only 
positive real coefficients, and so its value is given by Stage
\ref{stage4}.

\begin{stage}
If $\ket{\psi}=\sum_i\alpha_i\ket{\lambda_i}$, then
$\mc{V}(\ket{\psi})=\sum_i|\alpha_i|^2 x_i$.
\end{stage}

This last stage of the proof is simple and will not be spelled out in
detail.  It proceeds in exactly the same way as the proof of Stage
\ref{stage2}: any $n$-term measurement can be assembled by successive 
2-term measurements, using \textbf{Substitutivity}.

\section{Alternate form of Deutsch's proof}\label{altproof}

A slight change of Deutsch's assumptions allows us to simplify the theorem and
its proof. In this section we will be concerned with:
\begin{quote}
\textbf{Deutsch's Theorem (variant form):} If $\mc{V}$ is any value
function satisfying \textbf{Physicality}, \textbf{Dominance} and 
\textbf{Additivity}, it will be given by the Born Rule.
\end{quote}

The proof proceeds via part 3 of the
Equivalence Theorem (General Equivalence), which Deutsch did not use in his 
own proof. We define the \emph{expected utility} of a
\textbf{game} by $EU(\mc{G})=\sum_c W_\mc{G}(c) \,c$, where the sum ranges over the
distinct payoffs made.

As with Deutsch's own proof, we hold fixed the observable \op{X} 
to be measured, and suppose $\mc{P}(x)=\id{X}$ by default: this 
allows us to write
$\langle \ket{\psi} \rangle$ for $\langle \ket{\psi}, \op{X},
\mc{P}\rangle$.  In this case, we also write $EU(\ket{\psi})$ for 
$EU(\mc{G})$.

\begin{stageII}\label{stage2'}
If \mc{G} is an equally-weighted superposition of eigenstates of \op{X},
$\mc{V}(\ket{\psi})=EU(\psi)$.
\end{stageII}

Without loss of generality, suppose
$\ket{\psi}=(1/N)(\ket{\lambda_1}+ \cdots + \ket{\lambda_N}).$ Assume
first that all the $x_a$ are distinct,
let $\pi$ be an arbitrary permutation of $1, \ldots, N$, and define
$\mc{P}_\pi$ by $\mc{P}_\pi(x_a)=x_{\pi(a)}$.  Then by 
\textbf{Additivity},
\be \sum_\pi
\mc{V}(\ket{\psi},\op{X},\mc{P}_\pi)=\mc{V}(\ket{\psi},\op{X},\sum_\pi
\mc{P}_\pi)=(n-1)! \sum_i x_i\ee
since $\sum_\pi \mc{P}_\pi$ is just the constant payoff function that
gives a payoff of $(n-1)!(x_1+ \cdots +x_N)$ irrespective of the result
of the measurement.

But each of the $n!$ \textbf{games} $\langle \ket{\psi},\op{X},\mc{P}_\pi\rangle$ 
is a \textbf{game} in which each payoff $x_i$ occurs with weight $1/N$.  
Hence, by GE, all have equal value, and that value is just 
$\mc{V}(\ket{\psi})$.  Thus, $n! \mc{V}(\ket{\psi})=(n-1)!(x_1 + \cdots
+x_N)$, and the result follows.

If the $x_i$ are not all distinct, construct a sequence of operators
$\op{X}_m$ with eigenstates $x_{m,n}$ all distinct, so that
for each $n$ $\{x_{m,n}\}$ is an increasing sequence tending to $x_n$.
By \textbf{Dominance} this forces $\mc{V}(\ket{\psi},\op{X})\geq EU(\ket{\psi})$; 
repeating with a decreasing sequence proves the result.

\begin{stageII}\label{stage3'}
If $\ket{\psi}=\sum_i a_i \ket{\lambda_i}$, where the $a_i$ are all
rational, then $\mc{V}(\ket{\psi})=EU(\ket{\psi})$.
\end{stageII}

Any such state may be written 
\be \ket{\psi}=(1/\sqrt{N})\sum_i \sqrt{m_i}\ket{\lambda_i},\ee
where the $m_i$ are integers satisfying $\sum_i m_i=n$.  Such a game
associates a weight $m_i/N$ to payoff $x_i$.

But now consider an equally-weighted superposition \ket{\psi'} of $N$ 
eigenstates of
$\op{X}$ where a payoff of $x_1$ is given for any of the first $m_1$
eigenstates, $x_2$ for the next $m_2$, and so forth.  Such a game is
known (from stage \ref{stage2'}) to have value $(1/N)(m_1 x_1 + \cdots +
m_N x_n) \equiv EU(\ket{\psi})$.  But such a \textbf{game} also associates a weight
$m_i/N$ to payoffs of value $x_i$, so by GE we have
$\langle \ket{\psi} \rangle \simeq \langle \ket{\psi'}\rangle$ and the
result follows.

\begin{stageII}
For all states \ket{\psi} which are superpositions of finitely many 
eigenstates of \op{X},
$\mc{V}(\ket{\psi})=EU(\ket{\psi})$. 
\end{stageII}

By GE, it is sufficient to consider only states 
\be \ket{\psi}=\sum_i \alpha_i \ket{\lambda_i}\ee
with positive real
$\alpha_1$.  Let $\ket{\mu_i}$, $(1 \leq i \leq N)$, be a further set of
eigenstates of \op{X}, orthogonal to each other and to the
\ket{\lambda_i} and with eigenstates $y_i$ distinct from each other and
all strictly less than 
all of the $x_i$ (that we can always find such a set of states, or
reformulate the problem so that we can, is a consequence of GE).
For each $i$, $1\leq i\leq N$, let $a_i^n$ be an increasing
series of rational numbers converging on $\alpha_i^2$, and define
\be \ket{\psi_n}=\sum_i \sqrt{a_i^n}\ket{\lambda_i}+\sum_i \sqrt{\alpha^2_i-
a^n_i}\ket{\mu_i}.\ee

It follows from stage \ref{stage3'} that $\mc{V}(\ket{\psi_n})=EU(\ket{\psi_n})$, and
from \textbf{Dominance} that for all $n$, $\mc{V}(\ket{\psi}) \geq \mc{V}(\ket{\psi_n})$. 
Trivially $lim_{n\rightarrow\infty} EU(\ket{\psi_n})=EU(\ket{\psi})$, so
$\mc{V}(\ket{\psi})\geq EU(\ket{\psi})$.  Repeating the construction
with all the $y_i$ strictly greater than all the $x_i$ gives
$\mc{V}(\ket{\psi})\leq EU(\ket{\psi})$, and the result follows.
$\Box$

\section{Critique}\label{barnumI}

Barnum, Caves, Finkelstein, Fuchs and Schank, in their critique of Deutsch's paper 
\cite{BCFFS}, make three objections:
\begin{enumerate}
\item Deutsch claims to derive probability from the non-probabilistic
parts of quantum mechanics and decision theory. But the 
non-probabilistic part of decision theory already entails probability.
\item Deutsch's proof is technically flawed and contains a \emph{non
sequitur}.
\item Gleason's Theorem renders Deutsch's proof redundant.
\end{enumerate}

Responding on Deutsch's behalf to these objections provides a useful
analysis of the concepts and methods of his proof, and will be the topic of
this section.

We begin with Barnum \textit{et al}'s claim that the non-probabilistic part of 
decision theory already entails probabilities. They are referring results like the
`Probability representation theorem' quoted in section \ref{decision}, by
which we deduce that a rational agent confronted with uncertainty will always 
quantify that uncertainty by means of probabilities.\footnote{In fact,
they quote a related, but stronger result due to L.\,J.\,Savage, which
may be found in \citeN{savage} and is discussed in my
\citeyear{webversion}.}
 Since this theorem can
be proved with no reference to quantum theory (in particular, with no use of the
\textbf{Physicality} assumption), it certainly is not the case that Deutsch can 
claim to have derived
the very \emph{concept} of probability. (Of course, the representation
theorem certainly makes no mention of the Born rule; Deutsch \emph{can}
still claim to have derived the specific probability rule in question.)
In fact, Barnum \emph{et al}'s criticism can be sharpened: Deutsch cannot claim,
either, to have deduced the existence of \emph{uncertainty} from his starting-point,
for the decision-theoretic assumptions he makes apply only to a
situation where uncertainty is already present. (In the language of
section \ref{intro} this is to say that Deutsch's work arguably solves
the Quantitative Problem but not the Incoherence problem.)

What of Barnum \textit{et al}'s second criticism, of Deutsch's proof itself?
Translating their objections into my notation, their concern is
basically that Deutsch assumes, without justification, the rule
$\mc{V}(\ket{\psi},\op{X})=\mc{V}(\op{U}\ket{\psi},\op{U}\op{X}\opad{U})$
(this is, in effect, their equation (13), which they believe Deutsch
requires as an additional assumption). 

Of course, (13) is a direct consequence of \textbf{Physicality} (via
ME). The reason that this argument is unavailable to Barnum \emph{et al}
is that
they treat the measurement process as \emph{primitive}: to them (in this
paper at any rate) a measurement is \emph{axiomatically} specified by
the operator being measured, and consideration of the physical process
by which it is measured is irrelevant. 

This brings up an interesting ambiguity in the phrase 
``non-probabilistic part of quantum mechanics'', used in both papers.
Barnum \textit{et al} regard quantum mechanics in essentially the Dirac-von
Neumann paradigm: there are periods where the dynamics are unitary and
deterministic, followed by periods of stochastic evolution,
corresponding to measurements and where the probabilities are given by
the Born rule. In this framework, the ``non-probabilistic part''
naturally means the unitary, deterministic part, and the resulting theory is 
\emph{physically incomplete}, in the sense that it does not describe
even physically what happens during a measurement. This is the context
in which they are able to offer what is effectively an alternative
collapse rule (their equation (14)) which contradicts the Born rule.

To Deutsch, though, ``quantum mechanics'' means Everettian quantum
mechanics, which is (at least from a God's-eye view) a deterministic
theory. As such, to Deutsch the ``non-probabilistic part of quantum
mechanics'' means the \emph{whole} of quantum mechanics, and there is no
space for additional collapse rules --- but there is also no axiomatic
concept of measurement, hence the need for measurement neutrality to be
either assumed or argued for.

Understanding the difference between Deutsch's conception of QM, and
that of Barnum \textit{et al}, is also central to seeing why the latter
regard Gleason's Theorem as so central here.\footnote{Recall that Gleason's Theorem
tells us that for any map $f$ from the projectors on a Hilbert space of 3+
dimensions to $[0,1]$, such that if $\{\op{P}_i\}$ is an complete orthonormal set of
projectors then $\sum_i f(\op{P}_i)=1$, there exists some density
operator $\denop_f$ such that $f(\op{P})=\tr (\op{P}\denop_f)$.}
 For if we are looking for a
probabilistic rule to describe what happens at collapse, then Gleason's
Theorem tells us that this rule must be the Born Rule, provided only
that its probabilities are non-contextual. As an added bonus, it proves
that the physical state must be a (pure or mixed) Hilbert-space state,
which in principle allows the state to be regarded simply as an epistemic notion
(summarizing an agent's ignorance). 

The situation is rather different in the Everett interpretation. Here it
is the physical state that is our starting point, and the structure of a
measurement is derived rather than postulated. As such, there is no
logical space for a deduction of the state from the observables.

Nonetheless, might Gleason's Theorem provide us with the Born rule in
the Everett interpretation also? It could be used in the following
three-step proof of (the variant form of) Deutsch's theorem:
\begin{enumerate}
\item We begin by proving:
\begin{quote}
\textbf{Non-contextuality:} 
If \mc{V} is a value function satisfying \textbf{Dominance},
\textbf{Additivity} and \textbf{Physicality}, then
\be \mc{V}(\ket{\psi},\op{X},\mc{P})=\sum_{x\in \sigma(X)}
\mc{V}(\ket{\psi},\op{P}_X(x),\id{P_X(x)})\mc{P}(x).\ee
\end{quote}
(See the Appendix for the proof.)
\item Gleason's Theorem now tells us that
\be
\mc{V}(\ket{\psi},\op{P}_X(x),\id{P_X(x)})=\tr(\denop\op{P_X}(x)),\ee
where \denop\ is dependent on \ket{\psi} but not
on \op{X}. 
\item Let
\mc{S} be the one-dimensional Hilbert space spanned by \ket{\psi}, and
let \op{U} be the embedding map of \mc{S} into \mc{H}. Then
$\proj{\psi}{\psi}\op{U}=\op{U}\mathsf{1}$, and so 
\be \langle \ket{\psi},\mathsf{1},\id{1}\rangle \simeq 
\langle \ket{\psi},\proj{\psi}{\psi},\id{\proj{\psi}{\psi}}\rangle,\ee
by ME.
But since $\id{1}$ is a constant, the LHS game has value 1, hence so
does the RHS one. In turn this forces $\Pr(1)=1$, which is given only by
$\denop=\proj{\psi}{\psi}$.)
\end{enumerate}

Although this proof is valid,\footnote{It isn't valid in two-dimensional Hilbert 
spaces, of course --- but it would be disingenuous to claim this as an advantage 
for Deutsch's proof. That proof (and my variant on it) makes extensive use of auxiliary systems, of 
arbitrarily high dimension.} it certainly does not obviate the importance of 
Deutsch's proof:
\begin{enumerate}
\item Even applying probabilistic notions to branching requires decision
theory, to justify quantifying uncertainty by means of probability. (To
be sure, the use of decision theory to justify probabilities long predates Deutsch.)
\item The central insight in Deutsch's work (other than the observation that 
decision theory allows us to get clear exactly how probabilities apply
to the Everett interpretation) is that `games' do not correspond 
one-to-one with physical situations --- in my exegesis this is
represented by the Equivalence Theorem, of course. Steps 1 and 3 in the proof
above rely heavily on that theorem, without which Gleason's Theorem
falls short of establishing the Born rule.
\item More prosaically, the proof in this section is vastly more complex
than Deutsch's own proof. Proving
\textbf{Non-contextuality} from the decision-theoretic axioms is
scarcely simpler than proving Deutsch's theorem itself (since most of the hard
work goes into proving the Equivalence Theorem, which both utilize) and that is
before deploying Gleason's Theorem, the proof of which is far from
trivial. 
\end{enumerate}
Gleason's Theorem, then, seems to offer little or no
illumination of, or improvement to, Deutsch's proof.

\section{Measurement Neutrality}\label{neutrality}

We have seen that Deutsch's proof rests upon the observation that many \textbf{games}
--- \iec, triples $\langle \ket{\psi},\op{X},\mc{P}\rangle$ --- correspond 
to a single physical game. This is possible because we are treating measurement,
not as primitive, but as a physical process.

But this being so, there is a converse issue to address. Two \emph{different}
physical games can instantiate the same \textbf{game}; what of an agent who prefers
one to the other? Such an agent's preferences would not be represented 
effectively by a function \mc{V} on games.

Ruling this out is equivalent to assuming:
\begin{description}
\item[Measurement neutrality:] A rational agent is indifferent between two physical
games whenever they instantiate the same \textbf{game}.
\end{description}

The measurement neutrality assumption is hidden by Deutsch's (and my) notation.
In effect it is the assumption that, provided that
a given physical process fits the definition of a measurement of \op{X} on \ket{\psi}, the
details of how that measurement is done don't matter for decision-making
purposes. I will give two examples to show
why --- despite appearances --- it is not an altogether trivial assumption.

Firstly, observe that is the explanation
as to why Deutsch's theorem (which is, after all, a provable
\emph{theorem}) nonetheless has no implications for the probability
problem in  `hidden-variable' theories, 
such as the de Broglie-Bohm theory 
(\citeNP{bohm,holland}).
For in such theories, the physical state of a system is represented not just
by a Hilbert-space vector \ket{\psi}, but also by some set $\omega$ of hidden 
variables, so that the overall state is an ordered pair 
$\langle \ket{\psi}, \omega \rangle$. (In the de Broglie-Bohm
theory, for instance, $\omega$ is the position of the corpuscles.)
It is thus possible for two physical processes to agree as to the measurement carried
out, the payoff given, and the Hilbert-space state, but to disagree as to $\omega$ --- hence
a rational agent might prefer one process to the other.

To see how this might happen in practice, specialize to the 
de Broglie-Bohm theory, and to position measurements.
Suppose, in particular, that we consider a measurement of the spatial position of a 
particle in one dimension, and  assume that the quantum state is 
$\ket{\psi}=(1/\sqrt{2})(\ket{x}+\ket{-x})$, where $\ket{x}$ and $\ket{-x}$ 
are eigenvectors of position with eigenvalues $x$ and $-x$ respectively, and that
the payoff function is $\id{X}$. Stage 1 of Deutsch's proof (page \pageref{stage1}) 
establishes that the value of this game is zero, relying in the process
on the invariance of \ket{\psi} under reflection about the origin; but
unless the corpuscle state is also invariant about reflection, this
argument cannot be expected to apply to the de Broglie-Bohm theory.
And in fact, the corpuscle position \emph{cannot} be invariant under
reflection, except in conditions so extreme as to break the connection
between outcomes and the Hilbert-space state entirely, for the possible
outcomes of the measurement are $\pm x$ and so the corpuscle must have
one of those two positions.\footnote{We could, of course, try to get round this problem by considering a probability 
distribution over hidden variables
and requiring the \emph{distribution} to be symmetric. Fairly clearly, this forces a 
distribution assigning probability 0.5 to both $+x$ and $-x$. A Deutsch-style argument 
can now be applied, and yields the unedifying conclusion that \emph{if} the particle is 
at position $+x$ 50 \% of the time, it is rational to bet at even odds
that it will be found there when measured.}

Secondly, even in the context of the Everett interpretation measurement
neutrality rules out the strategy of regarding all branches as
equiprobable, independently of their amplitudes. For suppose I play a
game where I measure a spin-half particle and gain money if the result
is `spin-up' but lose money otherwise. Measurement device $\# 1$
(improbably) results in one branch for the spin-up result and one branch
for the spin-down result; device $\# 2$ incorporates a quantum random-number generator triggered
by a spin-up result, so that there are a trillion spin-up branches and only one spin-down one.
The equiprobability strategy tells me that I am as likely to gain as to
lose if I use device $\# 1$, but almost certain to win if I use device
$\# 2$ ---  yet measurement neutrality tells me that each is as good as
the other.

(To be sure, this particular result is \emph{already} implied if we adopt
Saunders' subjective-uncertainty (`SU') viewpoint on quantum-mechanical branching
(in which, recall, the correct attitude of an agent prior to branching
is to expect that they will experience one of the outcomes, but to be
uncertain as to which.) For device $\# 2$
is really just device $\# 1$, followed by  the triggering of the
randomizer, and so the SU description of its function is: ``either spin-up will occur, 
or spin-down. If it's
spin-up, some random process will occur in the innards of the measuring device (but it 
won't affect my winnings.)'' Looked at this way, the equiprobability assumption is 
already in trouble.)

The instinctive response to measurement neutrality, nonetheless, is usually that
it \emph{is} trivial --- who cares  exactly \emph{how} a
measurement device works, provided that it works?  What justifies
this instinctive response is presumably something like this:
let  A and B be possible
measurement devices for some observable $\op{X}$, and for each
eigenvalue $x$ of \op{X} let the agent be indifferent between the 
$x$-readout states of A and those of B.  Then if the agent is currently planning to use
device A, he can reason, ``Suppose I get an arbitrary result $x$.  Had I used device B I 
would still have got result $x$, and would not care about the difference caused in
the readout state by changing devices; therefore, I should be
indifferent about swapping to device B.''

The only problem with this account is that it assumes that this sort of
counterfactual reasoning is legitimate in the face of (subjective)
uncertainty, and this is at best questionable (see, \egc, \citeN{redhead} for
a discussion, albeit not in the context of the Everett interpretation).

For a defence secure against this objection, consider how the
traditional Dirac-von Neumann description of quantum mechanics treats
measurement.  In that account, a measurement device essentially does two
things.  When confronted with an eigenstate of the observable being
measured, it reliably evolves into a state which displays the associated
eigenvalue.  In addition, though, when confronted with a superposition
of eigenstates it causes wave-function collapse onto one of the
eigenstates (after which the device can be seen as reliably evolving
into a readout state, as above).  

In the Dirac-von Neumann description, it is rather mysterious why a
measurement device induces collapse of the wave-function.  One has the
impression that some mysterious power of the device, over and above its
properties as a reliable detector of eigenstates, induces the collapse,
and hence it is \textit{prima facie} possible that this power might
affect the probabilities of collapse (and thus that they might vary from
device to device) --- this would, of course, violate measurement neutrality.  
That this is not the case, and that the probabilities 
associated with the collapse are dependent only
upon the state which collapses (and indeed are equal to those stipulated
by the Born rule) is true by fiat in the Dirac-von Neumann description.

It is a strength of the Everett interpretation (at least as seen from
the SU viewpoint) that it recovers the subjective validity of the
Dirac-von Neumann description: once decoherence (and thus branching) occurs, 
subjectively there has been wave-function collapse.   Furthermore there is no
``mysterious power'' of the measurement device involved: measurement devices
by their nature amplify the superposition of eigenstates in the state to
be measured up to macroscopic levels, causing decoherence, and this in
turn leads to subjective collapse.

But this being the case, there is no rational justification for denying
measurement neutrality.  For the property of magnifying superpositions
to macroscopic scales is one which all measurement devices possess
equally, by definition --- so if this is the only property of the
devices relevant to collapse (after which the system is subjectively deterministic, and so 
differences between measurement devices are irrelevant) then no other
properties can be relevant to a rational allocation of probabilities.
The only relevant properties must be the state being measured, and the
particular superposition which is magnified to macroscopic scales ---
that is, the state being measured, and the observable being measured on it.

\section{Conclusion}\label{conclusion}

I have shown that Deutsch's approach does indeed allow a derivation of
the Born rule, from the following premises:
\begin{enumerate}
\item The correctness of the Everett interpretation.
\item The validity of regarding quantum branching, within the Everett
interpretation, as uncertain (at least subjectively).
\item A fairly strong set of decision-theoretic axioms.
\item Measurement neutrality.
\end{enumerate}
All four are needed. Without the Everett interpretation we cannot give
a realist description of QM which eschews hidden variables of any
sort, objectively stochastic dynamics, and an \emph{a priori} privileged
role for the observer. Without some license for agents to regard quantum branching 
as uncertain we cannot import classical decision theory into QM. Without
decision theory itself we have no license to transform uncertainty into
probability, and none of the constraints on those probabilities that
allow Deutsch's Theorem to be proven. And without measurement neutrality
we cannot draw any worthwhile conclusions from Deutsch's theorem, for it
is the assumption that connects the value function \mc{V} with real
decision-making.

All four are reasonable, however. The Everett interpretation's various
(non-probabilistic) foundational problems appear tractable; work by
Saunders, Vaidman and others seems to justify the application of
uncertainty-based concepts to branching; Deutsch's decision theory,
though based on quite strong axioms, seems perfectly reasonable for
small-scale betting;\footnote{In any case, Deutsch's decision theory can
be very substantially weakened; see \citeN{webversion}.} measurement
neutrality is at the least a plausible assumption, and may well be
defensible by either of the routes sketched out in section
\ref{neutrality}.

Deutsch's own conclusion claims that ``A decision maker who believes
only the non-probabilistic part of the theory, and is `rational' in the
sense defined by a strictly non-probabilistic restriction of classical
decision theory'' will make decisions according to the Born rule.''
Tacit in Deutsch's paper is that `the non-probabilistic part of the
theory' means no-collapse quantum mechanics, Everett-interpreted but
without prior assumptions about probability; it is less clear what the
`non-probabilistic restriction of classical decision theory' really
means, but if it simply means classical decision theory, shorn of
\emph{explicit} assumptions about probabilities and applied to branching
events as if they were uncertain-outcome events, then his claim seems
essentially correct. The implications for a satisfactory resolution of
the quantitative probability problem are then obvious --- and profoundly
important.

\section*{Acknowledgements}

For valuable discussions, I am indebted to Hannah Barlow, Katherine Brading, Harvey
Brown, Jeremy Butterfield, 
Adam Elga, Chris Fuchs, Hilary Greaves, Adrian Kent, Chris Timpson,
Wojciech Zurek, to all those at the 2002 Oxford-Princeton philosophy of
physics workshop, and especially to Simon Saunders and David Deutsch.  
Jeremy Butterfield and Simon Saunders also made detailed and helpful
comments on the longer version of this paper.

\section*{Appendix: proofs of the Probability Representation Theorem 
and Non-Contextuality}

\begin{quote}
\textbf{Lemma (Linearity):} If \mc{V} satisfies \textbf{Additivity} and 
\textbf{Dominance}, then for any sets of real numbers $\{a_i\}_{i=1}^N$ 
and payoffs $\{\mc{P}_i\}_{i=1}^N$, 
\be\mc{V}(\ket{\psi},\op{X},\sum_{i=1}^N a_i\mc{P}_i)=\sum_{i=1}^N a_i 
\mc{V}(\ket{\psi},\op{X},\mc{P}_i).\ee
\end{quote}
\textbf{Proof of lemma:} We will suppress \op{X} 
and \ket{\psi}, writing just \mc{V}(\mc{P}) for \mc{V}(\ket{\psi},\op{X},\mc{P}). 
Let $a$ be any positive real number and let $\{k_n\}$ and $\{m_n\}$ be
sequences of integers such that $\{k_m/m_n\}$ is an increasing sequence
tending to $a$. By \textbf{Dominance} and \textbf{Additivity} we have 
$ m_n \mc{V}(a \mc{P})\geq k_n \mc{V}(\mc{P})$
for all $n$, and hence $ \mc{V}(a \mc{P})\geq a \mc{V}(\mc{P})$.
Repeating this with a decreasing sequence, we get
$\mc{V}(a\mc{P})=a\mc{V}(\mc{P})$ for any $a\geq 0$; the extension to 
negative $a$ is trivial (just use \textbf{Zero-sum}) 
and the full result follows from \textbf{Additivity}.
$\Box$

\textbf{Proof of representation theorem:} For any $x\in\sigma(\op{X})$, define
$\delta_x(y)$ as equal to 1 when $y=x$, and equal to 0 otherwise.
Any payoff function \mc{P} for $\sigma(\op{X})$ can be expressed
uniquely as 
$ \mc{P}=\sum_{x\in\sigma(X)} \mc{P}(x) \delta_x,$
and so by \textbf{Linearity} we have
$ \mc{V}(\mc{P})=\sum_{x\in\sigma(X)} \mc{P}(x) \mc{V}(\delta_x);$
setting $\Pr(x)=\mc{V}(\delta_x)$ establishes (\ref{prob1}), and
putting $\mc{P}(x)=1$ for all $x$ gives (\ref{prob2}) as a special case.
$\Box$

\textbf{Proof of Non-contextuality:} By PE and \textbf{Linearity} we have 
\be \mc{V}(\mc{G})=
\sum_{x\in\sigma(X)} \mc{P}(x)\mc{V}(\ket{\psi},\op{X},\delta_x)
=\sum_{x\in\sigma(X)}\mc{P}(x)
\mc{V}(\ket{\psi},\delta_x(\op{X}),\id{\delta_x(X)})\ee
and the result is proved upon observing that $\delta_x(\op{X})=\op{P}_X(x)$.
$\Box$

\end{document}